\documentclass[12pt]{article}
\usepackage{amsmath}
\usepackage{bm}
\usepackage{color}

\begin{document}
\begin{center}
\begin{large}
{\bf Hydrogen atom in rotationally invariant noncommutative space}
\end{large}
\end{center}

\centerline {Kh. P. Gnatenko \footnote{E-Mail address: khrystyna.gnatenko@gmail.com, Tel:+380 32 2614443}, V. M. Tkachuk \footnote{E-Mail address: voltkachuk@gmail.com}}
\medskip
\centerline {\small \it Ivan Franko National University of Lviv, Department for Theoretical Physics,}
\centerline {\small \it 12 Drahomanov St., Lviv, 79005, Ukraine}

\abstract{We consider the noncommutative algebra which is
rotationally invariant. The hydrogen atom is studied in a
rotationally invariant noncommutative space. We find the
corrections to the energy levels of the hydrogen atom up to the
second order in the parameter of noncommutativity. The upper bound
of the parameter of noncommutativity is estimated on the basis of
the experimental results for $1s-2s$ transition frequency.


  PACS number(s): 03.65.-w, 02.40.Gh}
\section{Introduction}

The idea that space might have a noncommutative structure was proposed by Heisenberg and was formalized by Snyder \cite{Snyder}. In recent years, noncommutativity has met considerable interest due to development of String Theory \cite{Connes,Witten} and Quantum Gravity \cite{Doplicher}.

In the canonical version of noncommutative space the coordinate
and momentum operators satisfy the following commutation relations
 \begin{eqnarray}
[X_{i},X_{j}]=i\hbar\theta_{ij},\label{form101}\\{}
[X_{i},P_{j}]=i\hbar\delta_{i,j},\\{}
[P_{i},P_{j}]=0,\label{form10001}{}
\end{eqnarray}
where $\theta_{ij}$ is a constant antisymmetric matrix. Many
physical problems have been considered in this space (see, for instance, \cite{Gnatenko} and references therein). Among them the hydrogen atom was studied
\cite{Chaichian,Ho,Chaichian1,Chair,Stern,Zaim2,Adorno,Khodja}. In
\cite{Chaichian} the corrections to the energy levels of hydrogen atom were found up to the first order in the parameter of
noncommutativity. In this article the authors also obtained the
corrections to the Lamb shift within the noncommutative quantum
electrodynamic theory. In \cite{Ho} the hydrogen atom was studied
as a two-particle system in the case when particles of
opposite charges feel opposite noncommutativity. The quadratic
Stark effect was examined in \cite{Chair}. New result for shifts
in the spectrum of hydrogen atom in noncommutative space  was
presented in \cite{Stern}. In \cite{Zaim2} the hydrogen atom
energy levels were calculated in the framework of the
noncommutative Klein-Gordon equation. In \cite{Adorno, Khodja} the
Dirac equation with a Coulomb field was considered in
noncommutative space.

Hydrogen atom problem was also considered in the case of
space-time noncommutativity
\cite{Balachandran,Stern1,Moumni1,Moumni,Zaim}, space-space and
momentum-momentum noncommutativity \cite{Djemai, Kang,
Alavi}. Full phase-space noncommutativity in the Dirac equation was considered in \cite{Bertolami}. The authors concluded that in order to preserve gauge symmetry one should discard configuration space noncommutativity. In this article the hydrogen atom was studied. It was shown that only hyperfine structure is affected by the momentum noncommutativity. In the case of configuration space noncommutativity the issue of the gauge invariance is more complicated and was studied, for example, in \cite{Gitman}. In this article the authors prove the existence of generalized gauge transformations.

It is worth noting that in a three-dimensional noncommutative
space we face the problem of the rotational symmetry breaking
\cite{Chaichian,Sinha}. In order to preserve this symmetry new
classes of noncommutative algebras were explored. For instance, in
\cite{Moreno} the rotational invariance was preserved by foliating
the space with concentric fuzzy spheres. In \cite{Galikova} the
rotationally symmetric noncommutative space was constructed as a
sequence of fuzzy spheres. In this case the exact solution of the
hydrogen atom problem was found.  In \cite{Kupriyanov} the curved
noncommutative space was introduced to maintain the rotational
symmetry and the hydrogen atom spectrum was studied.

Also in order to preserve the rotation invariance the promotion of
the parameter of noncommutativity to an operator in Hilbert space
was suggested and the canonical conjugate momentum of this
operator was introduced \cite{Amorim}. In this case the author
considered an isotropic harmonic oscillator. The way to maintain
the $N$ dimensional rotation invariance was considered in
\cite{Bander}. In this article the coordinates are represented by
operators. The measured values of these operators are expectations
between generalized coherent states. In \cite{Bander1} the
noncommutative coordinates invariant under rotations were
presented. The author suggested identification of the coordinates
with the boost operators in SO(1,3).

Note, however, that in a two-dimensional space the rotational
symmetry survives even in the canonical version of
noncommutativity $[X,Y]=i\hbar\theta$, where $\theta$ is a
constant.

Much attention has also been received to studying of spherically
symmetric noncommutative spaces \cite{Buric, Murray, Buric1},
considering of the problem of violation of the Lorentz invariance (see
for example \cite{Carlson, Morita, Kase}). For instance, noncommutative gauge theory without Lorentz violation was proposed in
\cite{Carlson}. The authors considered a new class of noncommutative theories in which the parameter of noncommutativity is promoted to an antisymmetric tensor that
transforms as a Lorentz tensor. The generalizations of the
operator trace and star product were found. Therefore, the
Lagrangian was considered as a $\theta$-integrated quantity.

 In recent years, a new class of noncommutative theories that involve additional degrees of freedom namely spin degrees of freedom has been considered (see, for example, \cite{Falomir,Ferrari} and references therein). An important advantage of these theories is rotational invariance of noncommutative algebras. It is worth mentioning that in the case of spin-$1/2$ noncommutativity one is forced to extend the wave function to a two-component wave function.

In this article we consider the way of constructing the
rotationally invariant noncommutative space by the generalization
of the constant matrix $\theta_{ij}$ to a tensor. For this purpose we consider the idea to involve additional degrees of freedom. But in contrast to the spin noncommutativity we propose to involve some additional space coordinates. Namely, we propose to
construct the tensor of noncommutativity with the help of additional coordinates
governed by the harmonic oscillators. The corresponding algebra is
rotationally invariant. Extension of noncommutative algebra with the help of additional space coordinates allows us to describe quantum system by one-component wave function. The hydrogen atom is considered in the
rotationally invariant noncommutative space. We study the
perturbation of the hydrogen atom energy levels caused by the
noncommutativity of coordinates.

The article is organized as follows. In Section 2 the way to
preserve the rotational symmetry is considered. In Section 3 the
Hamiltonian of the hydrogen atom is studied in the rotationally
invariant noncommutative space. The corrections to the energy
levels of the hydrogen atom are found up to the second order in
the parameter of noncommutativity in Section 4. In addition, the
way to construct an effective Hamiltonian which is rotationally
invariant is proposed. Section 5 is devoted to calculation of the
corrections to the $ns$ levels. The upper bound of the parameter of noncommutativity is estimated in Section 6.
Conclusions are presented in Section 7.

\section{Rotationally invariant noncommutative space}

In order to preserve the rotational symmetry we propose to define the tensor of noncommutativity as follows
 \begin{eqnarray}
\theta_{ij}=\frac{\alpha}{\hbar}(a_{i}b_{j}-a_{j}b_{i}),\label{form130}
\end{eqnarray}
where $\alpha$ is a dimensionless constant and $a_{i}$, $b_{i}$ are
governed by a rotationally symmetric system. We suppose for
simplicity that $a_{i}$, $b_{i}$ are governed by the harmonic
oscillator
 \begin{eqnarray}
 H_{osc}=\frac{(p^{a})^{2}}{2m}+\frac{(p^{b})^{2}}{2m}+\frac{m\omega^{2} a^{2}}{2}+\frac{m\omega^{2}b^{2}}{2}.\label{form104}
 \end{eqnarray}
Note that harmonic oscillator has two independent units of measurement, namely the unit of length $\sqrt{\frac{\hbar}{m\omega}}$ and the  unit of energy $\hbar\omega$. It is generally believed that the parameter of noncommutativity of coordinates is of the order of the Planck scale, therefore we put
\begin{eqnarray}
\sqrt{\frac{\hbar}{m\omega}}=l_{p},\label{form200}
 \end{eqnarray}
 where $l_{p}$ is the Planck length. Independently of (\ref{form200}) we also consider the limit $\omega\rightarrow\infty$. In this case the distance between the energy levels of harmonic oscillator tends to infinity. Therefore, harmonic oscillator put into the ground state remains in it.

So, we propose to consider the following commutation relations
\begin{eqnarray}
[X_{i},X_{j}]=i\alpha(a_{i}b_{j}-a_{j}b_{i}),\label{form131}\\{}
[X_{i},P_{j}]=i\hbar\delta_{i,j},\\{}
[P_{i},P_{j}]=0.\label{form13331}{}
\end{eqnarray}

The coordinates $a_{i}$, $b_{i}$, and momenta $p^{a}_{i}$, $p^{b}_{i}$ satisfy the ordinary commutation relations $[a_{i},a_{j}]=0$, $[a_{i},p^{a}_{j}]=i\hbar\delta_{i,j}$, $[b_{i},b_{j}]=0$, $[b_{i},p^{b}_{j}]=i\hbar\delta_{i,j}$, also $[a_{i},b_{j}]=[a_{i},p^{b}_{j}]=[b_{i},p^{a}_{j}]=[p^{a}_{i},p^{b}_{j}]=0$. It is worth noting that $a_{i}$, $b_{i}$ commute with $X_{i}$ and $P_{i}$ and therefore $\theta_{ij}$ given by (\ref{form130}) commutes with $X_{i}$ and $P_{i}$ too. So,  $X_{i}$, $P_{i}$ and $\theta_{ij}$ satisfy the same commutation relations as in the case of the canonical version of noncommutativity. Therefore, in this sense algebra (\ref{form131})-(\ref{form13331}) is equivalent to (\ref{form101})-(\ref{form10001}). Note that in the case of the spin noncommutativity the situation is different. Namely, the tensor of noncommutativity is connected with the spin variable and does not commute with coordinates (see for instance \cite{Falomir}).

It is convenient to use the following representation
\begin{eqnarray}
X_{i}=x_{i}-\frac{\alpha}{2\hbar}\sum_{j}(a_{i}b_{j}-a_{j}b_{i})p_{j},
\label{form1}\\
P_{i}=p_{i},\label{form11111}
\end{eqnarray}
where the coordinates $x_{i}$ and momenta $p_{i}$ satisfy the ordinary commutation relations $[x_{i},x_{j}]=0$,
$[x_{i},p_{j}]=i\hbar\delta_{i,j}$. Taking into account (\ref{form1}),  it is clear that
\begin{eqnarray}
[X_{i},p^{a}_{j}]=\frac{i\alpha}{2}\left(b_{i}p_{j}-\delta_{i,j}({\bf b}\cdot{\bf p})\right),\label{form12345}\\{}
[X_{i},p^{b}_{j}]=-\frac{i\alpha}{2}\left(a_{i}p_{j}-\delta_{i,j}({\bf a}\cdot{\bf p})\right),\label{form123}\\{}
[P_{i},p^{a}_{j}]=[P_{i},p^{b}_{j}]=0.
\end{eqnarray}

 It is easy to show that noncommutative coordinates can be represented in a more convenient form
\begin{eqnarray}
X_{i}=x_{i}+\frac{1}{2}[{\bm{\theta}}\times{\bf p}]_{i},\label{form10}
\end{eqnarray}
where
\begin{eqnarray}
 {\bm{\theta}}=\frac{\alpha}{\hbar}[{\bf a}\times{\bf b}].\label{form1234}
\end{eqnarray}

So, we can represent noncommutative coordinates $X_{i}$ and
momenta $P_{i}$ by the coordinates $x_{i}$ and momenta $p_{i}$
which satisfy the ordinary commutation relations. Therefore the
Jacobi identity is satisfied and this can be easily checked for
all possible triplets of operators.

 Algebra (\ref{form131})-(\ref{form13331}) is manifestly rotationally invariant. It is clear that commutation relation (\ref{form131}) remains the same after rotation $X_{i}^{\prime}=U(\varphi)X_{i}U^{+}(\varphi)$, $a_{i}^{\prime}=U(\varphi)a_{i}U^{+}(\varphi)$,  $b_{i}^{\prime}=U(\varphi)b_{i}U^{+}(\varphi)$
\begin{eqnarray}
[X_{i}^{\prime},X_{j}^{\prime}]=i\alpha(a_{i}^{\prime}b_{j}^{\prime}-a_{j}^{\prime}b_{i}^{\prime}).
\end{eqnarray}
 where the rotation operator reads $U(\varphi)=e^{\frac{i}{\hbar}\varphi({\bf n}\cdot{\bf\tilde{L}})}$. Here ${\bf\tilde{L}}$ is the total angular momentum which we define as follows
 \begin{eqnarray}
 {\bf\tilde{L}}=[{\bf r}\times{\bf p}]+[{\bf a}\times{\bf p}^{a}]+[{\bf b}\times{\bf p}^{b}],
 \label{form500}
 \end{eqnarray}
 or taking into account (\ref{form11111}), (\ref{form10})
 \begin{eqnarray}
 {\bf\tilde{L}}=[{\bf R}\times{\bf P}]+\frac{1}{2}[{\bf P}\times[{\bm{\theta}}\times{\bf P}]]+[{\bf a}\times{\bf p}^{a}]+[{\bf b}\times{\bf p}^{b}],
 \end{eqnarray}
 where ${\bf r}=(x_{1},x_{2},x_{3})$, and ${\bf R}=(X_{1},X_{2},X_{3})$.

  Note that ${\bf\tilde{L}}$ satisfies the following commutation relations
  \begin{eqnarray}
 [X_{i},\tilde{L}_{j}]=i\hbar\varepsilon_{ijk}X_{k},\\{}
 [P_{i},\tilde{L}_{j}]=i\hbar\varepsilon_{ijk}P_{k},\\{}
  [a_{i},\tilde{L}_{j}]=i\hbar\varepsilon_{ijk}a_{k},\\{}
  [p^{a}_{i},\tilde{L}_{j}]=i\hbar\varepsilon_{ijk}p^{a}_{k},\\{}
  [b_{i},\tilde{L}_{j}]=i\hbar\varepsilon_{ijk}b_{k},\\{}
  [p^{b}_{i},\tilde{L}_{j}]=i\hbar\varepsilon_{ijk}p^{b}_{k},
  \end{eqnarray}
  which are the same as in ordinary space.
It is also worth mentioning that the total angular momentum (\ref{form500}) commutes with $r^{2}$, $a^{2}$, and $b^{2}$. Besides, it is easy to check that ${\bf\tilde{L}}$ commutes with the scalar products
\begin{eqnarray}
[\tilde{L}_{i},({\bf a}\cdot{\bf p})]=[\tilde{L}_{i},({\bf a}\cdot{\bf p})]=[\tilde{L}_{i},({\bf a}\cdot{\bf b})]=[\tilde{L}_{i},({\bf r}\cdot{\bf a})]=[\tilde{L}_{i},({\bf r}\cdot{\bf b})]=0.
\end{eqnarray}
As a consequence, it is clear that ${\bf\tilde{L}}$ commutes with the operator of distance which can be written in the following form
\begin{eqnarray}
R=\sqrt{\sum_{i}X_{i}^{2}}=\sqrt{\left({\bf
r}+\frac{1}{2}[{\bm{\theta}}\times{\bf
p}]\right)^{2}}=\sqrt{\left({\bf r}-\frac{\alpha }{2\hbar}{\bf
a}({\bf b}\cdot{\bf p})+\frac{\alpha }{2\hbar}{\bf b}({\bf
a}\cdot{\bf p})\right)^{2}}.
\end{eqnarray}
So, the distance remains the same after rotation $R^{\prime}=U(\varphi)R U^{+}(\varphi)=R$.

It is worth mentioning that there is also another way of generalization of the tensor of noncommutativity which gives the possibility to preserve the rotational symmetry
\begin{eqnarray}
\theta_{ij}=\frac{\alpha}{\hbar} l_{p}^{2} \varepsilon_{ijk} \tilde{a}_{k},\label{form132}
\end{eqnarray}
here $\tilde{a}_{k}=a_{k}/l_{p}$ are the dimensionless
coordinates corresponding to the harmonic oscillator and $\alpha$
is a constant.

Note that in both cases (\ref{form130}) and (\ref{form132}) in the limit $\alpha\rightarrow0$ we obtain the ordinary commutation relations.

At the end of this section we would like to discuss the physical meaning of additional coordinates $a_{i}$, $b_{i}$ which form the tensor of noncommutativity. We can treat them as some internal coordinates of particle. Quantum fluctuations of these coordinates lead effectively to a non-point-like particle, size of which is of the order of the Planck scale.

\section{Hamiltonian of the hydrogen atom}

In this section we study the Hamiltonian of the hydrogen atom in
the rotationally invariant noncommutative space
(\ref{form131})-(\ref{form13331}). We assume that in
noncommutative space Hamiltonian has a similar form as in the
ordinary space with commutative coordinates. So, the Hamiltonian
of the hydrogen atom reads
 \begin{eqnarray}
 H_{h}=\frac{P^{2}}{2M}-\frac{e^{2}}{R},
 \label{form888}
  \end{eqnarray}
where $R=\sqrt{\sum_{i}X_{i}^{2}}$. Besides, in rotationally
invariant noncommutative space we have to take into account
additional terms that correspond to the harmonic oscillator
(\ref{form104}). Therefore, we consider the total Hamiltonian as
follows
 \begin{eqnarray}
H=H_{h}+H_{osc}.\label{form13}
 \end{eqnarray}
 Note that $H_{h}$ does not commute with $H_{osc}$ because of commutation relations (\ref{form12345}), (\ref{form123}).

Let us expand  Hamiltonian (\ref{form13}) in the series over
${\bm{\theta}}$.  First, let us find the expansion for the
distance $R$ up to the second order in ${\bm{\theta}}$. Using
(\ref{form10}), we can write
\begin{eqnarray}
R=\sqrt{({\bf r}+\frac{1}{2}[{\bm{\theta}}\times{\bf p}])^{2}}=\sqrt{r^{2}-({\bm{\theta}}\cdot{\bf L})+\frac{1}{4}[{\bm{\theta}}\times{\bf p}]^{2}},\label{form140}
\end{eqnarray}
where $r=\sqrt{\sum_{i}x_{i}^{2}}$, ${\bf L}=[{\bf r}\times{\bf p}]$.
  Note, that the operators under the square root are noncommuting.  Nevertheless it is possible to find the expansion of $R$ over ${\bm{\theta}}$ in the similar form to the form which we have in the case of commuting operators but with the additional term $\theta^{2}f({\bf r})$ with unknown function $f({\bf r})$.
 \begin{eqnarray}
 R=r-\frac{1}{2r}({\bm{\theta}}\cdot{\bf L})-\frac{1}{8r^{3}}({\bm{\theta}}\cdot{\bf L})^{2}+\frac{1}{16}\left(\frac{1}{r}[{\bm{\theta}}\times{\bf p}]^{2}+[{\bm{\theta}}\times{\bf p}]^{2}\frac{1}{r}+\theta^{2}f({\bf r})\right).\label{form2}
 \end{eqnarray}

In order to find  $f({\bf r})$ let us square left- and right-hand sides of equation (\ref{form2}). In the second order over ${\bm{\theta}}$ we obtain
\begin{eqnarray}
r^{2}-({\bm{\theta}}\cdot{\bf L})+\frac{1}{4}[{\bm{\theta}}\times{\bf p}]^{2}=r^{2}-({\bm{\theta}}\cdot{\bf L})+\frac{1}{16}\left(2[{\bm{\theta}}\times{\bf p}]^{2}+r[{\bm{\theta}}\times{\bf p}]^{2}\frac{1}{r}+\frac{1}{r}[{\bm{\theta}}\times{\bf p}]^{2}r+2r\theta^{2}f({\bf r})\right).\label{form90}
\end{eqnarray}
Simplifying (\ref{form90}), we can write
\begin{eqnarray}
\frac{\hbar^{2}}{r^{4}}[{\bm{\theta}}\times{\bf r}]^{2}-r \theta^{2}f({\bf r})=0.\label{form91}
\end{eqnarray}
Finally, from equation (\ref{form91}) we find
\begin{eqnarray}
\theta^{2}f({\bf r})=\frac{\hbar^{2}}{r^{5}}[{\bm{\theta}}\times{\bf r}]^{2}.\label{form93}
\end{eqnarray}
Consequently, taking into account (\ref{form93}), we obtain the following expansion for the distance
  \begin{eqnarray}
 R=r-\frac{1}{2r}({\bm{\theta}}\cdot{\bf L})-\frac{1}{8r^{3}}({\bm{\theta}}\cdot{\bf L})^{2}+\frac{1}{16}\left(\frac{1}{r}[{\bm{\theta}}\times{\bf p}]^{2}+[{\bm{\theta}}\times{\bf p}]^{2}\frac{1}{r}+\frac{\hbar^{2}}{r^{5}}[{\bm{\theta}}\times{\bf r}]^{2}\right).\label{form12}
  \end{eqnarray}

 Now it is straightforward to expand the inverse distance $R^{-1}$ in the series over  ${\bm{\theta}}$.  Using (\ref{form12}), we find
\begin{eqnarray}
\frac{1}{R}=\frac{1}{r}+\frac{1}{2r^{3}}({\bm{\theta}}\cdot{\bf L})+\frac{3}{8r^{5}}({\bm{\theta}}\cdot{\bf L})^{2}-
\frac{1}{16}\left(\frac{1}{r^{2}}[{\bm{\theta}}\times{\bf p}]^{2}\frac{1}{r}+\frac{1}{r}[{\bm{\theta}}\times{\bf p}]^{2}\frac{1}{r^{2}}+\frac{\hbar^{2}}{r^{7}}[{\bm{\theta}}\times{\bf r}]^{2}\right)\label{form9}
\end{eqnarray}

So, taking into account (\ref{form9}), we can rewrite Hamiltonian (\ref{form13}) in the following form
\begin{eqnarray}
H=H_{0}+V,\label{form41}
\end{eqnarray}
where
\begin{eqnarray}
H_{0}=H_{h}^{(0)}+H_{osc},\label{form9999}
\end{eqnarray}
here $H_{h}^{(0)}=\frac{p^{2}}{2M}-\frac{e^{2}}{r}$ is the Hamiltonian of hydrogen atom in commutative space which commutes with $H_{osc}$, and $V$ is the perturbation caused by the noncommutativity of coordinates
\begin{eqnarray}
V=-\frac{e^{2}}{2r^{3}}({\bm{\theta}}\cdot{\bf L})-\frac{3e^{2}}{8r^{5}}({\bm{\theta}}\cdot{\bf L})^{2}+
\frac{e^{2}}{16}\left(\frac{1}{r^{2}}[{\bm{\theta}}\times{\bf p}]^{2}\frac{1}{r}+\frac{1}{r}[{\bm {\theta}}\times{\bf p}]^{2}\frac{1}{r^{2}}+\frac{\hbar^{2}}{r^{7}}[{\bm{\theta}}\times{\bf r}]^{2}\right).
\end{eqnarray}
It is worth emphasizing that ${\bm{\theta}}$ is defined by the coordinates $a_{i}$, $b_{i}$ by virtue of (\ref{form1234}). So, we have interaction terms in Hamiltonian $H$ that depend on the new coordinates $a_{i}$, $b_{i}$.

\section{Perturbation of the energy levels}

In order to find the corrections to the energy levels of the hydrogen atom let us use the perturbation theory.

 Note that $H_{h}^{(0)}$ commutes with $H_{osc}$ therefore the eigenvalues and the eigenstates of the unperturbed Hamiltonian (\ref{form9999}) read
\begin{eqnarray}
E^{(0)}_{n,\{n^{a}\},\{n^{b}\}}=-\frac{e^{2}}{2a_{B}n^{2}}+\hbar\omega(n_{1}^{a}+n_{2}^{a}+n_{3}^{a}+n_{1}^{b}+n_{2}^{b}+n_{3}^{b}+3),\\
\psi^{(0)}_{n,l,m,\{n^{a}\},\{n^{b}\}}=\psi_{n,l,m}\psi^{a}_{n_{1}^{a},n_{2}^{a},n_{3}^{a}}\psi^{b}_{n_{1}^{b},n_{2}^{b},n_{3}^{b}},
\end{eqnarray}
where $\psi_{n,l,m}$ are well known eigenfunctions of the hydrogen atom in ordinary space,  $\psi^{a}_{n_{1}^{a},n_{2}^{a},n_{3}^{a}}$, $\psi^{b}_{n_{1}^{b},n_{2}^{b},n_{3}^{b}}$ are the eigenfunctions of the three-dimensional harmonic oscillators, and $a_{B}$ is the Bohr radius.

Let us find the corrections to the energy levels of the hydrogen
atom in the case when the oscillators are in the ground state.
According to the perturbation theory, in the first order in $V$ we
have
\begin{eqnarray}
\Delta E^{(1)}_{n,l}=\langle\psi^{(0)}_{n,l,m,\{0\},\{0\}}|V|\psi^{(0)}_{n,l,m,\{0\},\{0\}}\rangle=\nonumber\\=\left\langle\psi^{(0)}_{n,l,m,\{0\},\{0\}}\left|-\frac{e^{2}}{2r^{3}}({\bm{\theta}}\cdot{\bf L})-\frac{3e^{2}}{8r^{5}}({\bm{\theta}}\cdot{\bf L})^{2}\right|\psi^{(0)}_{n,l,m,\{0\},\{0\}}\right\rangle+\nonumber\\+\frac{e^{2}}{16}
\left\langle\psi^{(0)}_{n,l,m,\{0\},\{0\}}\left|\frac{1}{r^{2}}[{\bm {\theta}}\times{\bf p}]^{2}\frac{1}{r}+\frac{1}{r}[{\bm {\theta}}\times{\bf p}]^{2}\frac{1}{r^{2}}+\frac{\hbar^{2}}{r^{7}}[{\bm{\theta}}\times{\bf r}]^{2}\right|\psi^{(0)}_{n,l,m,\{0\},\{0\}}\right\rangle.
\label{form999}
\end{eqnarray}

It is clear that
$\langle\psi^{a}_{0,0,0}\psi^{b}_{0,0,0}|\theta_{i}|\psi^{a}_{0,0,0}\psi^{b}_{0,0,0}\rangle=0$.
Therefore, in the first order in  ${\bm{\theta}}$  the corrections
to the energy levels vanish
\begin{eqnarray}
\left\langle\psi^{(0)}_{n,l,m,\{0\},\{0\}}\left|\frac{e^{2}}{2r^{3}}({\bm{\theta}}\cdot{\bf L})\right|\psi^{(0)}_{n,l,m,\{0\},\{0\}}\right\rangle=0.\label{form401}
\end{eqnarray}

In order to calculate the correction caused by the second term $\frac{3e^{2}}{8r^{5}}({\bm{\theta}}\cdot{\bf L})^{2}$ we use the following result (see for instance \cite{Wen-Chao})
\begin{eqnarray}
\left\langle\psi_{n,l,m}\left|\frac{1}{r^{5}}\right|\psi_{n,l,m}\right\rangle=\frac{4(5n^{2}-3l(l+1)+1)}{a_{B}^{5}n^{5}l(l+1)(l+2)(2l+1)(2l+3)(l-1)(2l-1)}.\label{form21}
\end{eqnarray}
It is easy to show that
\begin{eqnarray}
\langle\psi^{a}_{0,0,0}\psi^{b}_{0,0,0}|\theta_{i}\theta_{j}|\psi^{a}_{0,0,0}\psi^{b}_{0,0,0}\rangle=\frac{1}{2}\left(\frac{\alpha}{m\omega}\right)^{2}\delta_{i,j}=\frac{1}{3}\langle\theta^{2}\rangle\delta_{i,j},\label{form20}
\end{eqnarray}
where we use the notation
\begin{eqnarray}
\langle\theta^{2}\rangle=\langle\psi^{a}_{0,0,0}\psi^{b}_{0,0,0}|\theta^{2}|\psi^{a}_{0,0,0}\psi^{b}_{0,0,0}\rangle=\frac{3}{2}\left(\frac{\alpha}{m\omega}\right)^{2}=\frac{3\alpha^{2}l_{p}^{4}}{2\hbar^{2}}.\label{form887}
\end{eqnarray}
So, taking into account (\ref{form21}), (\ref{form20}), we find
\begin{eqnarray}
\left\langle\psi^{(0)}_{n,l,m,\{0\},\{0\}}\left|\frac{3e^{2}}{8r^{5}}({\bm{\theta}}\cdot{\bf L})^{2}\right|\psi^{(0)}_{n,l,m,\{0\},\{0\}}\right\rangle=\frac{\hbar^{2}e^{2}(5n^{2}-3l(l+1)+1)\langle\theta^{2}\rangle}{2a_{B}^{5}n^{5}(l+2)(2l+1)(2l+3)(l-1)(2l-1)}.\label{form402}
\end{eqnarray}
 In order to calculate the last term in (\ref{form999}) it is convenient to rewrite it as follows
 \begin{eqnarray}
  \frac{1}{r^{2}}[{\bm {\theta}}\times{\bf p}]^{2}\frac{1}{r}+\frac{1}{r}[{\bm {\theta}}\times{\bf p}]^{2}\frac{1}{r^{2}}+\frac{\hbar^{2}}{r^{7}}[{\bm{\theta}}\times{\bf r}]^{2}=\theta^{2}\frac{1}{r^{2}}p^{2}\frac{1}{r}+\theta^{2}\frac{1}{r}p^{2}\frac{1}{r^{2}}+\theta^{2}\frac{\hbar^{2}}{r^{5}} -\nonumber\\-\frac{1}{r^{2}}({\bm{\theta}}\cdot{\bf p})^{2}\frac{1}{r}-\frac{1}{r}({\bm {\theta}}\cdot{\bf p})^{2}\frac{1}{r^{2}}-\frac{\hbar^{2}}{r^{7}}({\bm{\theta}}\cdot{\bf r})^{2}.
 \end{eqnarray}
 Then taking into account (\ref{form20}), we obtain
  \begin{eqnarray}
\left\langle\psi^{a}_{0,0,0}\psi^{b}_{0,0,0}\left|\frac{1}{r^{2}}({\bm {\theta}}\cdot{\bf p})^{2}\frac{1}{r}+\frac{1}{r}({\bm {\theta}}\cdot{\bf p})^{2}\frac{1}{r^{2}}+\frac{\hbar^{2}}{r^{7}}({\bm{\theta}}\cdot{\bf r})^{2}\right|\psi^{a}_{0,0,0}\psi^{b}_{0,0,0}\right\rangle=\nonumber\\=
\frac{1}{3}\left(\frac{1}{r^{2}}p^{2}\frac{1}{r}+\frac{1}{r}p^{2}\frac{1}{r^{2}}+\frac{\hbar^{2}}{r^{5}}\right)\langle\theta^{2}\rangle.
\end{eqnarray}
As a result, the last term in (\ref{form999}) simplifies to
 \begin{eqnarray}
\left\langle\psi^{(0)}_{n,l,m,\{0\},\{0\}}\left|\frac{1}{r^{2}}[{\bm
{\theta}}\times{\bf p}]^{2}\frac{1}{r}+\frac{1}{r}[{\bm
{\theta}}\times{\bf
p}]^{2}\frac{1}{r^{2}}+\frac{\hbar^{2}}{r^{7}}[{\bm{\theta}}\times{\bf
r}]^{2}\right|\psi^{(0)}_{n,l,m,\{0\},\{0\}}\right\rangle=\nonumber\\=
\frac{2}{3}\left\langle\psi_{n,l,m}\left|\frac{1}{r^{2}}p^{2}\frac{1}{r}+\frac{1}{r}p^{2}\frac{1}{r^{2}}+\frac{\hbar^{2}}{r^{5}}\right|\psi_{n,l,m}\right\rangle\langle\theta^{2}\rangle.
\end{eqnarray}
Let us rewrite $ \frac{1}{r^{2}}p^{2}\frac{1}{r}+\frac{1}{r}p^{2}\frac{1}{r^{2}}+\frac{\hbar^{2}}{r^{5}}$ in the following form
 \begin{eqnarray}
 \frac{1}{r^{2}}p^{2}\frac{1}{r}+\frac{1}{r}p^{2}\frac{1}{r^{2}}+\frac{\hbar^{2}}{r^{5}}=\frac{1}{r^{3}}p^{2}+p^{2}\frac{1}{r^{3}}+\frac{5\hbar^{2}}{r^{5}}.
 \end{eqnarray}
 Therefore, we can write
 \begin{eqnarray}
\left\langle\psi_{n,l,m}\left|\frac{1}{r^{2}}p^{2}\frac{1}{r}+\frac{1}{r}p^{2}\frac{1}{r^{2}}+\frac{\hbar^{2}}{r^{5}}\right|\psi_{n,l,m}\right\rangle=-\frac{2\hbar^{2}}{a_{B}^{2}n^{2}}\left\langle\psi_{n,l,m}\left|\frac{1}{r^{3}}\right|\psi_{n,l,m}\right\rangle+\nonumber\\+\frac{4\hbar^{2}}{a_{B}}\left\langle\psi_{n,l,m}\left|\frac{1}{r^{4}}\right|\psi_{n,l,m}\right\rangle+5\hbar^{2}\left\langle\psi_{n,l,m}\left|\frac{1}{r^{5}}\right|\psi_{n,l,m}\right\rangle.\label{form400}
 \end{eqnarray}
 Using the following results (see for instance \cite{Wen-Chao})
\begin{eqnarray}
\left\langle\psi_{n,l,m}\left|\frac{1}{r^{3}}\right|\psi_{n,l,m}\right\rangle=\frac{2}{a_{B}^{3}n^{3}l(l+1)(2l+1)},\\
\left\langle\psi_{n,l,m}\left|\frac{1}{r^{4}}\right|\psi_{n,l,m}\right\rangle=\frac{4(3n^{2}-l(l+1))}{a_{B}^{4}n^{5}l(l+1)(2l+1)(2l+3)(2l-1)},
\end{eqnarray}
we can easy obtain an explicit expression for the last term in (\ref{form999}).

Finally in the first order of the perturbation theory we find the following corrections
\begin{eqnarray}
\Delta E^{(1)}_{n,l}=-\frac{\hbar^{2}e^{2}\langle\theta^{2}\rangle}{a_{B}^{5}n^{5}}\left(\frac{1}{6l(l+1)(2l+1)}-\frac{6n^{2}-2l(l+1)}{3l(l+1)(2l+1)(2l+3)(2l-1)}\right.+\nonumber\\\left.+\frac{5n^{2}-3l(l+1)+1}{2(l+2)(2l+1)(2l+3)(l-1)(2l-1)}-\frac{5}{6}\frac{5n^{2}-3l(l+1)+1}{l(l+1)(l+2)(2l+1)(2l+3)(l-1)(2l-1)}\right).\label{form411}
\end{eqnarray}

 In the second order of the perturbation theory we have
\begin{eqnarray}
\Delta
E_{n,l,m,\{0\},\{0\}}^{(2)}=\sum_{n^{\prime},l^{\prime},m^{\prime},\{n^{a}\},\{n^{b}\}}\frac{\left|\left\langle\psi^{(0)}_{n^{\prime},l^{\prime},m^{\prime},\{n^{a}\},\{n^{b}\}}\left|
V\right|\psi^{(0)}_{n,l,m,\{0\},\{0\}}\right\rangle\right|^{2}}{E^{(0)}_{n}-E^{(0)}_{n^{\prime}}-\hbar\omega(n^{a}_{1}+n^{a}_{2}+n^{a}_{3}+n^{b}_{1}+n^{b}_{2}+n^{b}_{3})},\label{form311}
\end{eqnarray}
where the set of numbers $n^{\prime}$, $l^{\prime}$, $m^{\prime}$,
$\{n^{a}\}$, $\{n^{b}\}$ does not coincide with the set $n$, $l$,
$m$, $\{0\}$, $\{0\}$, and
$E^{(0)}_{n}=-e^{2}/(2a_{B}n^{2})$ is the unperturbed energy
of the hydrogen atom.
 It is worth noting that because of our assumption (\ref{form200}) matrix elements $\left\langle\psi^{(0)}_{n^{\prime},l^{\prime},m^{\prime},\{n^{a}\},\{n^{b}\}}\left| V\right|\psi^{(0)}_{n,l,m,\{0\},\{0\}}\right\rangle$ do not depend on $\omega$.  Therefore, in the case of $\omega\rightarrow\infty$ we obtain
\begin{eqnarray}
\mathop{lim}\limits_{\omega\rightarrow\infty}\Delta
E_{n,l,m,\{0\},\{0\}}^{(2)}=0.\label{form300}
\end{eqnarray}

Consequently, taking into account (\ref{form411}), (\ref{form300}), the corrections to the hydrogen atom energy levels up to the second order in parameter of noncommutativity read
\begin{eqnarray}
\Delta E_{n,l}=\Delta E^{(1)}_{n,l}.\label{form22}
\end{eqnarray}

At the end of this section we would like to note that one can obtain the same correction to the energy spectrum defining an effective Hamiltonian.
As was mentioned above, in the case of $\omega\rightarrow\infty$ the harmonic oscillator is always in the ground state. Therefore, we construct an effective Hamiltonian in the following form
\begin{eqnarray}
H^{eff}=\langle\psi^{a}_{0,0,0}\psi^{b}_{0,0,0}|H|\psi^{a}_{0,0,0}\psi^{b}_{0,0,0}\rangle,
\end{eqnarray}
 Taking into account (\ref{form888}), (\ref{form9}), (\ref{form20}), we obtain
\begin{eqnarray}
H_{h}^{eff}=\langle\psi^{a}_{0,0,0}\psi^{b}_{0,0,0}|H_{h}|\psi^{a}_{0,0,0}\psi^{b}_{0,0,0}\rangle=\frac{ p^{2}}{2M}-\frac{e^{2}}{r}-\frac{e^{2}L^{2}}{8r^{5}}\langle\theta^{2}\rangle+\nonumber\\ +\frac{e^{2}}{24}\left(\frac{1}{r^{2}}p^{2}\frac{1}{r}+\frac{1}{r}p^{2}\frac{1}{r^{2}}+\frac{\hbar^{2}}{r^{5}}\right)\langle\theta^{2}\rangle.\label{form1001}
\end{eqnarray}
It is important to note that effective Hamiltonian (\ref{form1001}) is rotationally invariant.
Using the first order perturbation theory, we find the same corrections to the energy levels as (\ref{form22}).

It is worth mentioning that in the case of $l=0$ or $l=1$
corrections  (\ref{form411}) are divergent. The problem of
divergence of the corrections also appears in the deformed space
with minimal length. In order to overcome this problem the
modified perturbation theory was proposed \cite{Tkachuk, Stetsko}.

In this article in order to estimate the upper bound of the parameter of noncommutativity we are interested in the corrections to the $ns$ levels.  In next section we propose the way to find the corrections to the $ns$ levels of the hydrogen atom in the rotationally invariant noncommutative space. We plan to consider the problem of divergence of the corrections in the case of $l=1$  in a forthcoming publication.

\section{Corrections to the $ns$ levels of the hydrogen atom}

 In order to find the corrections to the $ns$ levels let us rewrite the perturbation caused by the noncommutativity of coordinates in the following form
  \begin{eqnarray}
 V=-\frac{e^{2}}{R}+\frac{e^{2}}{r}=-\frac{e^{2}}{\sqrt{r^{2}-({\bm{\theta}}\cdot{\bf L})+\frac{1}{4}[{\bm{\theta}}\times{\bf p}]^{2}}}+\frac{e^{2}}{r}.\label{form600}
  \end{eqnarray}

 Using the perturbation theory, and taking into account (\ref{form300}), the corrections to the $ns$ levels read
\begin{eqnarray}
\Delta E_{ns}=\left\langle\psi^{(0)}_{n,0,0,\{0\},\{0\}}\left|\frac{e^{2}}{r}-\frac{e^{2}}{\sqrt{r^{2}-({\bm{\theta}}\cdot{\bf L})+\frac{1}{4}[{\bm{\theta}}\times{\bf p}]^{2}}}\right|\psi^{(0)}_{n,0,0,\{0\},\{0\}}\right\rangle.
 \label{form721}
 \end{eqnarray}
Note that $({\bm{\theta}}\cdot{\bf L})$ commutes with $[{\bm{\theta}}\times{\bf p}]^{2}$ and $r^{2}$. Also  it is clear that $({\bm{\theta}}\cdot{\bf L})\psi^{(0)}_{n,0,0,\{0\},\{0\}}=0$ because $\psi^{(0)}_{n,0,0,\{0\},\{0\}}$ does not depend on angles. Therefore, we can rewrite (\ref{form721}) in the following form
\begin{eqnarray}
\Delta E_{ns}=\left\langle\psi^{(0)}_{n,0,0,\{0\},\{0\}}\left|\frac{e^{2}}{r}-\frac{e^{2}}{\sqrt{r^{2}+\frac{1}{4}[{\bm{\theta}}\times{\bf p}]^{2}}}\right|\psi^{(0)}_{n,0,0,\{0\},\{0\}}\right\rangle.
 \end{eqnarray}
  Using (\ref{form20}), up to the second order in the parameter of noncommutativity we can write
 \begin{eqnarray}
\Delta E_{ns}=\left\langle\psi_{n,0,0}\left|\frac{e^{2}}{r}-\frac{e^{2}}{\sqrt{r^{2}+\frac{1}{6}\langle\theta^{2}\rangle p^{2}}}\right|\psi_{n,0,0}\right\rangle=\left\langle R_{n,0}\left|\frac{e^{2}}{r}-\frac{e^{2}}{\sqrt{r^{2}+\frac{1}{6}\langle\theta^{2}\rangle p_{r}^{2}}}\right|R_{n,0}\right\rangle,
 \label{form700}
 \end{eqnarray}
where $p_{r}=-i\hbar\frac{1}{r}\frac{\partial}{\partial r}r$, and $R_{n,0}=\sqrt{\frac{4}{a_{B}^{3}n^{5}}}e^{-\frac{r}{na_{B}}}L_{n-1}^{1}\left(\frac{2r}{a_{B}n}\right)$ is the radial wavefunction,  $L^{1}_{n-1}\left(\frac{2r}{a_{B}n}\right)$ are the generalized Laguerre polynomials.

 First, let us find the correction to the $1s$ energy level of the hydrogen atom. It is convenient to introduce the dimensionless coordinate  $\rho=r\left(\frac{6}{\hbar^{2}\langle\theta^{2}\rangle}\right)^{\frac{1}{4}}$. Therefore, we have
\begin{eqnarray}
\Delta E_{1s}=\left\langle R_{1,0}\left|\frac{e^{2}}{r}-\frac{e^{2}}{\sqrt{r^{2}+\frac{1}{6}\langle\theta^{2}\rangle p_{r}^{2}}}\right|R_{1,0}\right\rangle=\frac{4e^{2}\beta^{2}}{a_{B}}\int_{0}^{\infty}d\rho\rho^{2}e^{-\beta\rho}\left(\frac{1}{\rho}-\frac{1}{\sqrt{\rho^{2}+ p_{\rho}^{2}}}\right)e^{-\beta\rho}.
 \label{form730}
 \end{eqnarray}
 where  $p_{\rho}=-i\frac{1}{\rho}\frac{\partial}{\partial \rho}\rho$ and $\beta=\left(\frac{\hbar^{2}\langle\theta^{2}\rangle}{6a_{B}^{4}}\right)^{\frac{1}{4}}$.

  In order to calculate (\ref{form730}) we propose to expand  $e^{-\beta\rho}$
over the eigenfunctions of $\rho^{2}+p_{\rho}^{2}$ which we denote $\phi_{k}$
 \begin{eqnarray}
 e^{-\beta\rho}=\sum_{k=0}^{\infty}C_{k}\phi_{k}.
  \label{form720}
  \end{eqnarray}

The eigenfunctions and the eigenvalues of $\rho^{2}+p_{\rho}^{2}$ are as follows (see for instance \cite{Yanez})
 \begin{eqnarray}
 \phi_{k}=\sqrt{\frac{2k!}{\Gamma(k+\frac{3}{2})}}e^{-\frac{\rho^{2}}{2}}L^{\frac{1}{2}}_{k}(\rho^{2}),\label{form740}\\
 \lambda_{k}=2\left(2k+\frac{3}{2}\right).
  \end{eqnarray}

So, using (\ref{form740}), from (\ref{form720}) we find

   \begin{eqnarray}
   C_{k}=\sqrt{\frac{2 k!}{\Gamma(k+\frac{3}{2})}}\int_{0}^{\infty}d\rho\rho^2e^{-\frac{\rho^{2}}{2}-\beta \rho}L^{\frac{1}{2}}_{k}\left(\rho^{2}\right).
  \end{eqnarray}

As a result, we can write
 \begin{eqnarray}
 \int_{0}^{\infty}d\rho\rho^{2}e^{-\beta\rho}\frac{1}{\sqrt{\rho^{2}+ p_{\rho}^{2}}}e^{-\beta\rho}= \sum_{k=0}^{\infty}\frac{C_{k}^{2}}{\sqrt{\lambda_{k}}}.
 \label{form711}
 \end{eqnarray}
Similarly for the first term in (\ref{form730}) we have
 \begin{eqnarray}
\int_{0}^{\infty}d\rho\rho e^{-2\beta\rho}= \sum_{k=0}^{\infty}C_{k}\int_{0}^{\infty}d\rho \rho e^{-\beta\rho}\phi_{k}=\sum_{k=0}^{\infty}C_{k}I_{k},
 \label{form710}
 \end{eqnarray}
 where
 \begin{eqnarray}
 I_{k}=\sqrt{ \frac{2k!}{\Gamma(k+\frac{3}{2})}}\int_{0}^{\infty}d\rho\rho e^{-\frac{\rho^{2}}{2}-
 \beta \rho}L^{\frac{1}{2}}_{k}\left(\rho^{2}\right).
\end{eqnarray}
 So, taking into account (\ref{form711}), (\ref{form710}), the correction to the $1s$ energy level reads
  \begin{eqnarray}
\Delta
E_{1s}=\frac{4e^{2}\beta^{2}}{a_{B}}\sum_{k=0}^{\infty}\left(C_{k}I_{k}-\frac{C_{k}^{2}}{\sqrt{\lambda_{k}}}\right)=\frac{e^{2}\beta^{2}}{a_{B}}S_{1s}(\beta),
 \end{eqnarray}
 where we use the notation
 \begin{eqnarray}
  S_{1s}(\beta)=4\int_{0}^{\infty}d\rho\rho^{2}e^{-\beta\rho}\left(\frac{1}{\rho}-\frac{1}{\sqrt{\rho^{2}+ p_{\rho}^{2}}}\right)e^{-\beta\rho}=4\sum_{k=0}^{\infty}\left(C_{k}I_{k}-\frac{C_{k}^{2}}{\sqrt{\lambda_{k}}}\right).
 \end{eqnarray}
  It is worth noting that $S_{1s}(0)$ has a finite value
 \begin{eqnarray}
  S_{1s}(0)=16\sqrt{\frac{2}{\pi}}\sum_{k=0}^{\infty}\frac{\Gamma(k+\frac{3}{2})}{k!}\left({}_{2}F_{1}\left(-k,\frac{1}{2};\frac{3}{2};2\right)-\sqrt{\frac{\pi}{8k+6}}\right)=1.72006\ldots
 \end{eqnarray}
 where ${}_{2}F_{1}\left(-k,\frac{1}{2};\frac{3}{2};2\right)$ is the hypergeometric function. Therefore it is clear that the asymptotic of $\Delta E_{1s}$ for $\beta\rightarrow0$ ($\theta\rightarrow0$ ) is as follows
  \begin{eqnarray}
  \Delta E_{1s}=\frac{e^{2}\beta^{2}}{a_{B}}S_{1s}(0)=\frac{e^{2}}{a_{B}^{3}}\sqrt{\frac{\hbar^{2}\langle\theta^{2}\rangle}{6}}
S_{1s}(0).
  \end{eqnarray}

 Likewise, we can find the corrections to the excited $s$ levels
 \begin{eqnarray}
\Delta E_{ns}=\frac{e^{2}\beta^{2}}{a_{B}n^{5}}S_{ns}(\beta),
\end{eqnarray}
where
\begin{eqnarray} S_{ns}(\beta)=4\int_{0}^{\infty}d\rho\rho^{2}e^{-\frac{\beta\rho}{n}}L_{n-1}^{1}\left(\frac{2\beta\rho}{n}\right)\left(\frac{1}{\rho}-\frac{1}{\sqrt{\rho^{2}+ p_{\rho}^{2}}}\right)e^{-\frac{\beta\rho}{n}}L_{n-1}^{1}\left(\frac{2\beta\rho}{n}\right).
\end{eqnarray}
It is easy to show that
\begin{eqnarray}
S_{ns}(0)=S_{1s}(0)n^{2}\simeq1.72n^{2}.
\end{eqnarray}
As a result we obtain the following corrections to the $ns$ levels of the hydrogen atom
\begin{eqnarray}
  \Delta E_{ns}
 =\frac{e^{2}}{a^{3}_{B}n^{3}}\sqrt{\frac{\hbar^{2}\langle\theta^{2}\rangle}{6}}
 S_{1s}(0).
 \label{form714}
 \end{eqnarray}

 \section{Estimation of the upper bound of the parameter of noncommutativity}

Let us estimate the upper bound of the parameter of
noncommutativity. For this purpose we use the result of
measurement of the hydrogen $1s-2s$ transition frequency
\cite{Parthey}. The authors obtained $f_{1s-2s} = 2466061413187018
(11)$Hz with relative uncertainty of $4.5\times10^{-15}$.

Using (\ref{form714}), the correction to the energy of the $1s-2s$
transition reads
\begin{eqnarray}
\Delta_{1,2}=\Delta E_{2s}-\Delta E_{1s}=-\frac{7e^{2} S_{1s}(0) }{8\sqrt{6}a_{B}^{3}}\hbar\sqrt{\langle\theta^{2}\rangle},\\
\frac{\Delta_{1,2}}{E^{(0)}_{2}-E^{(0)}_{1}}=-\frac{7}{3\sqrt{6}}\frac{S_{1s}(0)}{a_{B}^{2}} \hbar\sqrt{\langle\theta^{2}\rangle},
\end{eqnarray}
where $E^{(0)}_{n}=-e^{2}/(2a_{B}n^{2})$ is the unperturbed energy of the hydrogen atom.
Assuming that $|\Delta_{1,2}|/(E^{(0)}_{2}-E^{(0)}_{1})$ does not
exceed $4.5\times10^{-15}$, we find
\begin{eqnarray}
\frac{7}{3\sqrt{6}}\frac{S_{1s}(0)}{a_{B}^{2}}\hbar\sqrt{\langle\theta^{2}\rangle}\leq4.5\times10^{-15},\\
\hbar
\sqrt{\langle\theta^{2}\rangle}\leq7.7\times10^{-36}\,\textrm{m}^{2}.\label{form55}
\end{eqnarray}
This assumption was also considered
in order to find the upper bound of the minimal length (see, for
instance, \cite{Brau,Quesne}).

 It is worth mentioning
that our result (\ref{form55}) is stronger than the upper bound
obtained on the basis of the data on the Lamb shift in
\cite{Chaichian}.

Taking into account (\ref{form887}), we can also estimate the value of $\alpha$ as follows
\begin{eqnarray}
\alpha\leq2.4\times10^{34}.
\end{eqnarray}

\section{Conclusion}

In this article we have considered an important problem of the
rotational symmetry breaking in noncommutative space. In order to
preserve this symmetry we proposed the generalization of the
tensor of noncommutativity (\ref{form130}) or (\ref{form132})
which gives the possibility to construct rotationally invariant
algebras.

The hydrogen atom has been examined in the rotationally invariant
noncommutative space. Using the perturbation theory, the
corrections to the hydrogen atom energy levels have been found. It
is worth noting that we have faced a problem of divergence of the
corrections to the $ns$ and $np$ energy levels. For estimation of
the upper bound of the parameter of noncommutativity we have used
the experimental result for $1s-2s$ transition frequency which is
measured with a high accuracy \cite{Parthey}. Therefore, the way
to find the corrections to the $ns$ levels has been proposed.
Comparing our results with the experimental data we have found
\begin{eqnarray}
\hbar
\sqrt{\langle\theta^{2}\rangle}\leq7.7\times10^{-36}\,\textrm{m}^{2}.
\end{eqnarray}
This result is stronger than the upper bound estimated on the
basis of the data on the Lamb shift in \cite{Chaichian}.

In addition, an effective rotationally invariant Hamiltonian has
been constructed. We have defined the effective Hamiltonian of the
hydrogen atom and obtained the corrections to the energy levels.
It is worth noting that these results are in agreement with
corrections (\ref{form22}) obtained from the total Hamiltonian
(\ref{form13}).

\section{Acknowledgements}
The authors are grateful to Yu. S. Krynytskyi for his advices and many useful comments. We also thank Dr. A. A. Rovenchak and Dr. M. M. Stetsko for a careful reading of the manuscript.


\begin{thebibliography}{99}
\bibitem{Snyder} H. Snyder, Phys. Rev. {\bf 71}, 38 (1947).
\bibitem{Connes} A. Connes, M.R. Douglas, A. Schwarz, J. High Energy Phys. {\bf 9802}, 003 (1998).
\bibitem{Witten} N. Seiberg, E. Witten, J. High Energy Phys. {\bf 9909}, 032 (1999).
\bibitem{Doplicher} S. Doplicher, K. Fredenhagen, J.E. Roberts, Phys. Lett. B {\bf 331}, 39 (1994).
\bibitem{Gnatenko}  Kh.P. Gnatenko Phys. Lett. A {\bf 377}, 3061 (2013).
\bibitem{Chaichian} M. Chaichian, M.M. Sheikh-Jabbari, A. Tureanu, Phys. Rev. Lett. {\bf 86}, 2716 (2001).
\bibitem{Ho} Pei-Ming Ho, Hsien-Chung Kao, Phys. Rev. Lett. {\bf 88}, 151602 (2002).
\bibitem{Chaichian1} M. Chaichian, M.M. Sheikh-Jabbari, A. Tureanu, Eur. Phys. J. C {\bf 36}, 251 (2004).
\bibitem{Chair} N. Chair, M.A. Dalabeeh, J. Phys. A: Math. Gen. {\bf 38}, 1553 (2005).
\bibitem{Stern} A. Stern, Phys. Rev. Lett. {\bf 100}, 061601 (2008).
\bibitem{Zaim2} S. Zaim, L. Khodja, Y. Delenda, Int. J. Mod. Phys. A  {\bf 26}, 4133 (2011).
\bibitem{Adorno} T.C. Adorno, M. C. Baldiotti, M. Chaichian, D.M. Gitman, A. Tureanu, Phys. Lett.  B {\bf 682}, 235 (2009).
\bibitem{Khodja} L. Khodja, S. Zaim, Int. J. Mod. Phys. A {\bf 27}, 1250100 (2012).



\bibitem{Balachandran} A.P. Balachandran, A. Pinzul Mod. Phys. Lett. A {\bf 20}, 2023 (2005).
\bibitem{Stern1} A. Stern, Phys. Rev. D {\bf 78}, 065006 (2008).
\bibitem{Moumni1} M. Moumni, A. BenSlama, S. Zaim, Journal of Geometry and Physics {\bf 61}, 151 (2011).
\bibitem{Moumni} M. Moumni, A. BenSlama, S. Zaim, The African Review of Physics {\bf 07}, 83 (2012).
  \bibitem{Zaim} S. Zaim, Y. Delenda, J. Phys.: Conf. Ser. {\bf 435}, 012020 (2013).


\bibitem{Djemai} A.E.F. Djemai, H. Smail, Commun. Theor. Phys. {\bf 41}, 837 (2004).
\bibitem{Kang} Li Kang, Chamoun Nidal, Chin. Phys. Lett. {\bf 23}, 1122 (2006).
\bibitem{Alavi} S.A. Alavi, Mod. Phys. Lett. A {\bf 22}, 377 (2007).
\bibitem{Bertolami} O. Bertolami, R. Queiroz, Phys. Lett. A {\bf 375}, 4116 (2011).
\bibitem{Gitman} D.M. Gitman, V.G. Kupriyanov J. Math. Phys. {\bf 51}, 022905 (2010).
\bibitem{Sinha} D. Sinha, B. Chakraborty, F.G. Scholtz, J. Phys. A: Math. Theor. {\bf 45}, 105308 (2012).

\bibitem{Moreno} E.F. Moreno, Phys. Rev. D {\bf 72}, 045001 (2005).
\bibitem{Galikova} V. G\'alikov\'a, P. Presnajder, J. Phys: Conf. Ser. {\bf 343}, 012096 (2012).
\bibitem{Kupriyanov} V.G. Kupriyanov, J. Phys. A: Math. Theor. {\bf 46}, 245303 (2013).
\bibitem{Amorim} R. Amorim, Phys. Rev. Lett. {\bf 101}, 081602 (2008).
\bibitem{Bander}  M. Bander, J. High Energy Phys. {\bf 03}, 040 (2006).
\bibitem{Bander1} M. Bander, Phys. Rev. D {\bf 75}, 105010 (2007).

\bibitem{Buric} M. Buri\'c, J. Madore, Eur. Phys. J. C {\bf 58},  347 (2008).
\bibitem{Murray} S. Murray, J. Govaerts, Phys. Rev. D {\bf 83},  025009 (2011).
\bibitem{Buric1} M. Buri\'c, J. Madore, Eur. Phys. J. C {\bf 74},  2820 (2014).

\bibitem{Carlson} C.E. Carlson, C.D. Carone, N. Zobin, Phys. Rev. D {\bf 66}, 075001 (2002).
\bibitem{Morita} K. Morita, Prog. Theor. Phys. {\bf 108}, 1099 (2003).
\bibitem{Kase} H. Kase, K. Morita, Y. Okumura, E. Umezawa, Prog. Theor. Phys. {\bf 109}, 663 (2003).


\bibitem{Falomir} H. Falomir, J. Gamboa, J. L\'opez-Sarri\'on, F. M\'endez, P.A.G. Pisani, Phys. Lett. B {\bf 680}, 384 (2009).
 \bibitem{Ferrari} A.F. Ferrari, M. Gomes, V.G. Kupriyanov, C.A. Stechhahn, Phys. Lett. B {\bf 718}, 1475 (2013).
\bibitem{Wen-Chao} Wen-Chao Qiang, Shi-Hai Dong, Phys. Scripta {\bf 70}, 276 (2004).
\bibitem{Tkachuk} M.M. Stetsko, V.M. Tkachuk, Phys. Rev. A {\bf 74}, 012101 (2006).
\bibitem{Stetsko} M.M. Stetsko, Phys. Rev. A {\bf 74}, 062105 (2006).
\bibitem{Yanez} R.J. Y\'a\~nez, W. Van Assche,  J. S. Dehesa Phys. Rev. A {\bf 50}, 3065 (1994).
\bibitem{Parthey} A. Matveev, C.G. Parthey, K. Predehl et al., Phys. Rev. Lett. {\bf 110}, 230801 (2013).
\bibitem{Brau} F. Brau, J. Phys. A: Math. Gen. {\bf 32}, 7691 (1999).
\bibitem{Quesne} C. Quesne, V.M. Tkachuk, Phys. Rev. A {\bf 81}, 012106
(2010).

\end{thebibliography}
\end{document}